# DESIGN OF A RECONFIGURABLE DSP PROCESSOR WITH BIT EFFICIENT RESIDUE NUMBER SYSTEM[1]

Chaitali Biswas Dutta[1], Partha Garai[2] and Amitabha Sinha[3]

[1] Research Scholar, Dept of CSE, University of Kalyani, India
Assistant Professor, Dept of Computer Application, Girijananda Chowdhury Institute of Management & Technology, Guwahati, India
`mail.chaitali@yahoo.in`
[2] Machine Intelligence Unit, Indian Statistical Institute, Kolkata, 203, BT Road, Kolkata - 700108, India
`parthagarai_r@isical.ac.in`
[3] School of Information Technology, West Bengal University of Technology, BF-142, Sector-1, Salt Lake City, Kolkata-700064, India
`amitabha@wbut.ac.in`

## ABSTRACT

*Residue Number System (RNS), which originates from the Chinese Remainder Theorem, offers a promising future in VLSI because of its carry-free operations in addition, subtraction and multiplication. This property of RNS is very helpful to reduce the complexity of calculation in many applications. A residue number system represents a large integer using a set of smaller integers, called residues. But the area overhead, cost and speed not only depend on this word length, but also the selection of moduli, which is a very crucial step for residue system. This parameter determines bit efficiency, area, frequency etc. In this paper a new moduli set selection technique is proposed to improve bit efficiency which can be used to construct a residue system for digital signal processing environment. Subsequently, it is theoretically proved and illustrated using examples, that the proposed solution gives better results than the schemes reported in the literature. The novelty of the architecture is shown by comparison the different schemes reported in the literature. Using the novel moduli set, a guideline for a Reconfigurable Processor is presented here that can process some predefined functions. As RNS minimizes the carry propagation, the scheme can be implemented in Real Time Signal Processing & other fields where high speed computations are required.*

## 1. INTRODUCTION

In recent times, Residue Number System (RNS) are being popular to implement a variety of specialized high-performance Digital Signal Processing (DSP) systems for its carry-free nature. Weighted number systems such as the binary number system, decimal number system etc has a carry chain [1]. It is often limiting the performance of arithmetic operations [2, 3]. In RNS, several residue digits represent a number. So, arithmetic operations like additions, subtractions and multiplications of higher bit numbers can be decomposed and performed in set of parallel sub-operation. As a result carry propagation, which is a genuine problem in weighted number systems, will be minimized in residue systems. RNS is extremely efficient for many applications

---

such as digital signal processing [4,5,6] communications engineering, computer security (cryptography) [6] etc.

Generally, number of bits required in residue number system is greater than that of weighted number systems because RNS gives the number of residues same as the cardinality of the moduli set, increasing the number of bit required to express it in RNS. A number system is said to have higher bit efficiency if the bit required to represent a particular dynamic range is lower. There are many important parameters that determine the efficiency of RNS and bit efficiency is one of them. The bit efficiency depends on the choice of the moduli set [7]. There are several techniques [7,8,9] for moduli set generation reported in the literature $\{2^n, 2^n+1, 2^n-1\}$, $\{2^n, 2^n-1, 2^{n-1}-1\}$ and $\{2^{2n}+1, 2^n+1, 2^n-1\}$. For these schemes no algorithm is given to generate a moduli set; they are generated heuristically by finding a suitable n. The contributions of paper are following:

1. proposed an algorithm to generate any moduli set with finite cardinality in a given dynamic range.
2. bit efficiency of the proposed scheme is better than all other scheme given in the literature.
3. theoretical analysis and proof of the proposed scheme to show that the proposed solution gives better results than the existing scheme [9].
4. Applicability of this scheme in a reconfigurable DSP Processor

## 2. BRIEF OVERVIEW OF RNS

RNS uses a set of numbers $(r_0, r_1, r_2, ..., r_{t-1})$, which is mapped with some number X in any other number system using a set of integers $m_0, m_1, m_2$ & $m_{t-1}$ called moduli. These numbers are relatively prime, that is, GCD $(m_i, m_j) = 1$ for $i \neq j$. Let X be a decimal number and N be the product of all moduli. N is called dynamic range. Then RNS can represent any numbers from 0 to (N-1). Now r = (X mod m) where r is the remainder of a number X with respect to modulus m. Number X will be represented by n-tuple $(r_0, r_1, r_2, ..., r_{t-1})$, where $r_i = (X \mod m_i)$ and $0 \le i \le t-1$ [7].

Now if X ≡ (r1x, r2x, ..., rmx) & Y ≡ (r1y, r2y, ..., rmy) then X ⊗ Y ≡ (r1x ⊗ r1y, r2x ⊗ r2y..., rmx ⊗ rmy)                    ... (1)

where ⊗ represents any arithmetic operator and can be like addition, subtraction or multiplication. So from the equation (1) it is clear that using RNS, integer arithmetic can be broken down into some independent parts which can be calculated in parallel fashion without a carry between each component. So the operations can be performed much faster even faster than the special hardwares like Carry Look Ahead Adder [2], Carry Select Adder [2], Carry Save Adder [1,2], and Wallace Tree Multiplier [1,2], Array Multiplier [2] etc. When the size of a modulus increases, it gives large reminders having multiple bits. So when an arithmetic operation is performed on those reminders, carries are propagated within the small range. These special hardwares mentioned above can be used to do the operations that help increasing the processing time.

The arithmetic operations are implemented with residue number system [3,10], depending on the choice of the moduli. The Chinese Remainder Theorem (CRT) [11,12,13] may rightly be viewed as one of the most important fundamental results in the theory of RNS. The CRT is useful for many other operations and above all it is very helpful in case of RNS to binary conversion [8,13]. Mainly New Chinese Remainder Theorem is introduced for this conversion [14,15,16,17] . CRT is assured that if the moduli of a RNS are chosen appropriately then each number in the dynamic range will have a unique representation in the residue system.

In the literature, there are a few kinds of moduli sets. A set of any given moduli is called a general-moduli set, as it is efficient for RNS systems with a large dynamic range. The three-moduli sets,

$S_{M_1} = \{2^n, 2^n+1, 2^n-1\}$, $\qquad$ $S_{M_2} = \{2n, 2n+1, 2n-1\}$, $\qquad$ $S_{M_3} = \{2^n, 2^n-1, 2^{n-1}-1\}$ and

$S_{M_4} = \{2^{2n}+1, 2^n+1, 2^n-1\}$ are four cases of the general-moduli sets and these sets are widely used for residue number system with a medium dynamic range [9,18].

In this paper we propose an algorithm that generates moduli sets for medium to large dynamic ranges. The scheme also attempts to keep the number of bits required to represent the moduli to a minimum.

## 3. SCHEME FOR IMPROVING BIT EFFICIENCY

The bits required to implement all the blocks of RNS number are depends on moduli set. Let N be the number of bit then $2^N$ is called dynamic range. Now $(r_0, r_1, r_2, ..., r_{t-1})$ denotes the t-moduli set, where $r_0, r_1, r_2, ..., r_{t-1}$ all are relatively prime and product of these t numbers should be greater or equal to $2^N-1$. Total bits required is calculated as $\lceil \log_2 r_0 \rceil + \lceil \log_2 r_1 \rceil + \lceil \log_2 r_2 \rceil + ... + \lceil \log_2 r_{t-1} \rceil$. Bit width of the different arithmetic block (like, adder, multiplier) of residue systems depend on the number $\lceil \log_2 r_0 \rceil + \lceil \log_2 r_1 \rceil + \lceil \log_2 r_2 \rceil + ... + \lceil \log_2 r_{t-1} \rceil$. Lower the value of this term, more optimized design of RNS in terms of bit width in achieved. Choice can be made over the various moduli set (like, three-moduli, four- moduli) and also the number within the set.

In this section we describe an algorithm to generate any number of moduli set for a given precision.

### Module find_moduli(N,n,SM)

//Input: N (no. of Bit), n  (no. of moduli set)
//Output: SM   (Efficient moduli set)

Step 1: $x = \left\lceil \sqrt[n]{2^N-1} \right\rceil$

Step 2:  if x is even then $2n = x$

$\qquad$ else $2n = x+1$

Step 3:  When n = 3

$\qquad$ if $((2n)(2n+1)(2n-1) \geq (2^N-1))$ then $S_M = \{2n, 2n+1, 2n-1\}$

$\qquad$ else   n will be incremented till $((2n)(2n+1)(2n-1) \geq (2^N-1))$ condition will be satisfied.

Step 4:  if n = 4 then

$\qquad$ Let $k = \left\lceil 2^N-1 \Big/ \{(2n)(2n+1)(2n-1)\} \right\rceil$ Find the smallest number $k_1 \geq k$ , where $k_1$ is relatively prime to $2n$ , $2n+1$ and $2n-1$ .

$\qquad$ $S_M = \{2n, 2n+1, 2n-1, k1\}$

… …      … … … … … … … … …

… …      … … … … … … … … …

Step p:  if n = p then

$$k = \left\lceil 2^N - 1 \middle/ \{(2n)(2n+1)(2n-1)\} \right\rceil$$

$$k^1 = \left\lceil \sqrt[(p-3)]{k} \right\rceil$$

Find the smallest number $k_1 \geq k^1$, where $k_1$ is relatively prime to $2n$, $2n+1$ and $2n-1$.

Therefore, $S_M = \{2n, 2n+1, 2n-1, k_1\}$

Again,

$$k = \left\lceil 2^N - 1 \middle/ \{(2n)(2n+1)(2n-1)(k_1)\} \right\rceil$$

$$k^2 = \left\lceil \sqrt[(p-4)]{k} \right\rceil$$

Find the smallest number $k_2 \geq k^2$, where $k_2$ is relatively prime to $2n$, $2n+1$, $2n-1$ and $k_1$.

Therefore, $S_M = \{2n, 2n+1, 2n-1, k_1, k_2\}$

… … … … … … … … … … … … … …
… … … … … … … … … … … … … …

$$k = \left\lceil 2^N - 1 \middle/ \{(2n)(2n+1)(2n-1)(k_1)...(k_{p-4})\} \right\rceil \qquad k^{p-3} = \left\lceil \sqrt[(p-(p-1))]{k} \right\rceil$$

Find the smallest number $k_{p-3} \geq k^{p-3}$, where $k_{p-3}$ is relatively prime to $2n$, $2n+1$ and $2n-1$, $k_1, \ldots, k_{p-4}$.

Therefore, $S_M = \{2n, 2n+1, 2n-1, k_1, k_2, k_3, \ldots k_{p-3}\}$

## Theorem

The bit efficiency of the present scheme is better than the existing scheme of linear complexity.

## Proof

We will first proof the result for three moduli set, and then extend the result for the general case.

Given a three moduli set $\{h_1, h_2, h_3\}$ and another three moduli set $\{h_1', h_2', h_3'\}$. Bit count of $\{h_1, h_2, h_3\}$ $\left(i.e., \lceil \log(h_1) \rceil + \lceil \log(h_2) \rceil + \lceil \log(h_3) \rceil \right)$ is better than that of $\{h_1', h_2', h_3'\}$ $\left(i.e., \lceil \log(h_1') \rceil + \lceil \log(h_2') \rceil + \lceil \log(h_3') \rceil \right)$ iff $(h_1 + h_2 + h_3) \geq (h_1' + h_2' + h_3')$.

Now we consider two types of three-moduli set, one for our proposed scheme $(i.e, \{2n', 2n'+1, 2n'-1\})$ and another for [9] $(i.e, \{2^{n_1}, 2^{n_1}+1, 2^{n_1}-1\})$.

As $2^n$ can always as an even number, i.e., 2n but the reverse does not hold. $(2n' + (2n'+1) + (2n'-1)) \leq (2^{n_1} + (2^{n_1}+1) + (2^{n_1}-1))$;

They are equal when 2n can be represented as $2^n$, less than otherwise.

For the general case, we start with the four moduli set. Now we consider two types of four-moduli set, one for our proposed scheme $(i.e, \{2n', 2n'+1, 2n'-1, k\})$ and another for [9] $(i.e, \{2^{n_1}, 2^{n_1}+1, 2^{n_1}-1, 2^{n_2} \pm 1\})$.

We have already shown that $(2n'+(2n'+1)+(2n'-1)) \leq (2^{n_1}+(2^{n_1}+1)+(2^{n_1}-1))$. Now, from the construction of the moduli set [9], $k$ is the smallest number relative prime to $2n', 2n'+1, 2n'-1$ and $2n' \times (2n'+1) \times (2n'-1) \times k \geq 2^N$. So, $\{2n'+(2n'+1)+(2n'-1)+k\} \leq \{2^{n_1}+(2^{n_1}+1)+(2^{n_1}-1)+(2^{n_2}+1)\}$.

This logic follows for any size moduli-set.

Now an example will be given to illustrate the algorithm:
Let N = 32

Therefore, dynamic range is 0 to $2^{32}-1$ i.e., 0 to 4294967295

Now we find the moduli sets $S_{M_i}$, for $i = 3, 4, 5, 6$.

For $n = 3$, we have $S_{M_3} = \{2n, 2n+1, 2n-1\}$ = $\{1626, 1627, 1625\}$

$$\left[ \begin{array}{l} \because x = \left\lceil \sqrt[n]{2^N - 1} \right\rceil = \left\lceil \sqrt[3]{2^{32}-1} \right\rceil = 1626 \\ and\ x\ is\ even\ then\ 2n = x = 1626 \\ and\ (2n)(2n+1)(2n-1) > (2^N - 1), \\ i.e., (1626)(1627)(1625) > (2^{32}-1) \end{array} \right]$$

For n = 4, we have $S_{M_4} = \{2n, 2n+1, 2n-1, k_1\}$ = $\{256, 257, 255, k_1\}$

$$\left[ \begin{array}{l} \because x = \left\lceil \sqrt[n]{2^N - 1} \right\rceil = \left\lceil \sqrt[4]{2^{32}-1} \right\rceil = 256 \\ and\ x\ is\ even\ then\ 2n = x = 256 \end{array} \right]$$

$k_1$ is calculated in the following way:
Here,

$$k = \left\lceil 2^N - 1 \Big/ \{(2n)(2n+1)(2n-1)\} \right\rceil = 257$$

Now we have to find the smallest number $k_1 \geq k$, where $k_1$ is relatively prime to $2n$, $2n+1$ and $2n-1$, i.e., $256, 257, 255$ Here, $k_1 = 259$. 259 is the smallest number where $259 > 257$ and 259 is also relatively prime to $256, 257, 255$. So, $S_{M_4} = \{256, 257, 255, 259\}$

For n = 5, we have
$S_{M_5} = \{2n, 2n+1, 2n-1, k_1, k_2\}$ = $\{86, 87, 85, k_1, k_2\}$

$$\left[ \begin{array}{l} \because x = \left\lceil \sqrt[n]{2^N - 1} \right\rceil = \left\lceil \sqrt[5]{2^{32}-1} \right\rceil = 85 \\ and\ x\ is\ odd\ then\ 2n = x+1 = 86 \end{array} \right]$$

$k_1$ and $k_2$ are calculated in the following way:

To find $k_1$,

$$k = \left\lceil 2^N - 1 \middle/ \{(2n)(2n+1)(2n-1)\} \right\rceil = 6754$$

$$k^1 = \left\lceil \sqrt[(p-3)]{k} \right\rceil = \left\lceil \sqrt{6754} \right\rceil = 83$$

Now we have to find the smallest number $k_1 \geq k^1$, where $k_1$ is relatively prime to $2n$, $2n+1$ and $2n-1$, i.e., $86, 87$ and $85$. Here, $k_1 = 89$. $89$ is the smallest number where $89 > 83$ and $89$ is also relatively prime to $86, 87$ and $85$.

To find $k_2$,

$$k = \left\lceil 2^N - 1 \middle/ \{(2n)(2n+1)(2n-1)(k_1)\} \right\rceil = 76$$

$$k^2 = \left\lceil \sqrt[(p-4)]{k} \right\rceil = 76$$

Now we have to find the smallest number $k_2 \geq k^2$, where $k_2$ is relatively prime to $2n, 2n+1, 2n-1$ and $k_1$, i.e., $86, 87, 85$ and $89$. Here, $k_2 = 77$. $77$ is the smallest number where $77 > 76$ and $77$ is also relatively prime to $86, 87, 85$ and $89$. So, $S_{M_5} = \{86, 87, 85, 89, 77\}$

When n = 6, we have
$$S_{M_6} = \{2n, 2n+1, 2n-1, k_1, k_2, k_3\} = \{42, 43, 41, k_1, k_2, k_3\}$$

$$\left[ \because x = \left\lceil \sqrt[5]{2^N - 1} \right\rceil = \left\lceil \sqrt[6]{2^{32} - 1} \right\rceil = 41 \right]$$
$$\left[ and \ x \ is \ odd \ then \ 2n = x+1 = 42 \right]$$

$k_1$, $k_2$ and $k_3$ is calculated as the following way:

To find $k_1$,

$$k = \left\lceil 2^N - 1 \middle/ \{(2n)(2n+1)(2n-1)\} \right\rceil = 58005$$

$$k^1 = \left\lceil \sqrt[(p-3)]{k} \right\rceil = \left\lceil \sqrt[3]{58005} \right\rceil = 39$$

Now we have to find the smallest number $k_1 \geq k^1$, where $k_1$ is relatively prime to $2n$, $2n+1$ and $2n-1$, i.e., $42, 43$ and $41$.

Here, $k_1 = 47$. $47$ is the smallest number where $47 > 39$ and $47$ is also relatively prime to $42, 43$ and $41$.

To find $k_2$,

$$k = \left\lceil 2^N - 1 \middle/ \{(2n)(2n+1)(2n-1)(k_1)\} \right\rceil = 1235$$

$$k^2 = \left\lceil \sqrt[(p-4)]{k} \right\rceil = 36$$

Now we have to find the smallest number $k_2 \geq k^2$, where $k_2$ is relatively prime to $2n$, $2n+1, 2n-1$ and $k_1$, i.e., $42, 43, 41$ and $47$. Here, $k_2 = 37$. $37$ is the smallest number where $37 > 36$ and $37$ is also relatively prime to $42, 43, 41$ and $47$.

To find $k_2$,

$$k = \left\lceil 2^N - 1 \middle/ \{(2n)(2n+1)(2n-1)(k_1)(k_2)\} \right\rceil = 34 \quad k^3 = \left\lceil \sqrt[(p-5)]{k} \right\rceil = 34$$

Now we have to find the smallest number $k_3 \geq k^3$, where $k_3$ is relatively prime to $2^n, 2n+1, 2n-1$, $k_1$ and $k_2$, i.e., $42, 43, 41, 47$ and $37$. Here, $k_3 = 53$. $53$ is the smallest number where $53 > 34$ and $53$ is also relatively prime to $42, 43, 41, 47$ and $37$.

So, $S_{M_6} = \{42, 43, 41, 47, 37, 53\}$.

## 4. COMPARISON OF BIT EFFICIENCY

In this section we will compute the number of bits required to implement the moduli set generated by the algorithm discussed in the last section. We also compare the bit efficiency of the moduli set proposed by us with some stander moduli set namely $S_{M_1} = (2^n, 2^n+1, 2^n-1)$, $S_{M_2} = (2^n, 2^n-1, 2^{n-1}-1)$ and $S_{M_3} = (2^{2n}+1, 2^n+1, 2^n-1)$.

Among them $S_{M_1} = (2^n, 2^n+1, 2^n-1)$ is the most standard and widely used. We compare bit efficiency of $S_{M_1} = (2^n, 2^n+1, 2^n-1)$ with that of $S_M = \{2n, 2n+1, 2n-1, k_1, k_2, k_3, \ldots k_{p-3}\}$, which is proposed by us, up to six-moduli set. We also present a comparison with other sets for three-moduli set. Bits required to implement the moduli set $S_M = \{n_1, n_2, n_3, \ldots, n_p\}$ are $\lceil \log_2 n_1 \rceil + \lceil \log_2 n_2 \rceil + \lceil \log_2 n_3 \rceil + \ldots + \lceil \log_2 n_p \rceil$

Table 1. Comparison of bit efficiency of proposed scheme for moduli set with cardinality 3, 4, 5, 6.

| N | | Proposed Scheme | |
|---|---|---|---|
| | | Moduli Set | No. of Bits |
| 16 | n=3 | (42,43,41) | 18 |
| | n=4 | (16,17,15,19) | 19 |
| | n=5 | (10,11,9,13,7) | 19 |
| | n=6 | (8,9,7,11,5,13) | 22 |
| 20 | n=3 | (102,103,101) | 21 |
| | n=4 | (32,33,31,35) | 23 |
| | n=5 | (16,17,15,19,23) | 24 |
| | n=6 | (12,13,11,17,7,19) | 25 |
| 32 | n=3 | (1626,1627,1625) | 33 |
| | n=4 | (256,257,255,259) | 35 |
| | n=5 | (86,87,85,89,77) | 35 |
| | n=6 | (42,43,41,47,37,53) | 36 |

Table 2. Comparison of bit efficiency of proposed scheme with standard approaches for three moduli set

| | Proposed Scheme | | $S_{M_1}=(2^n,2^n+1,2^n-1)$ | | $S_{M_2}=(2^n,2^n-1,2^{n-1}-1)$ | | $S_{M_3}=(2^{2n}+1,2^n+1,2^n-1)$ | |
|---|---|---|---|---|---|---|---|---|
| N | 3-Moduli Set | #Bits | 3-Moduli Set | #Bits | 3-Moduli Set | #Bits | 3-Moduli Set | #Bits |
| 6 | (6,7,5) | 9 | (8,9,7) | 11 | (8,7,3) | 9 | (17,5,3) | 10 |
| 10 | (12,13,11) | 12 | (16,17,15) | 14 | (16,15,7) | 12 | (65,9,7) | 14 |
| 16 | (42,43,41) | 18 | (64,65,63) | 20 | (64,63,31) | 18 | (257,17,15) | 18 |
| 24 | (256,257,255) | 26 | (512,513,511) | 29 | (512,511,255) | 27 | (4097,65,63) | 26 |
| 32 | (1626,1627,1625) | 33 | (2048,2049,2047) | 35 | (4096,4097,2047) | 37 | (65537,257,255) | 34 |

Fig 1: Graphical representation of Comparison of bit efficiency of proposed scheme with standard approaches for three moduli set

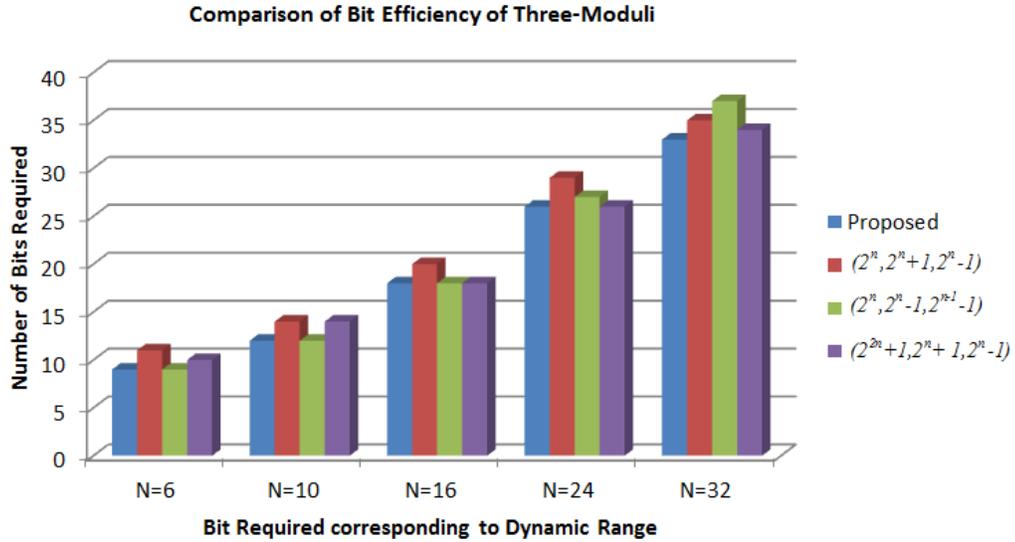

In Table 1 we present bits required for moduli set with cardinality three, four, five, six generated by the proposed approach. In [9], an excellent scheme is proposed where moduli set $M_G$ is generated as $\{2^{n_1}, 2^{n_1}+1, 2^{n_1}-1, 2^{n_2} \pm 1, ..., 2^{n_t} \pm 1\}$, where $n_1 > n_i$, and $n_1$ as well as $n_i$'s ($i = 2,3,...,t$) need to be chosen such that all these moduli are co-prime numbers. This scheme is implemented up to 6 moduli sets. It may be noted that [9] is most widely accepted and our proposed scheme gives better results in most of the cases, for the remaining cases, it gives the same bit efficiency as given by [9]. In Table 2 comparison has been done for three-moduli set with different schemes namely $S_{M_1} = (2^n, 2^n+1, 2^n-1)$, $S_{M_2} = (2^n, 2^n-1, 2^{n-1}-1)$ and $S_{M_3} = (2^{2n}+1, 2^n+1, 2^n-1)$. The results have been illustrated graphically in Figure 1.

From these tables it is observed that bits required in the proposed scheme are minimum than that of other schemes of order O(n). In other words our algorithm generates the most efficient moduli set.

## 5. GENERAL ARCHITECTURE OF RECONFIGURABLE RNS PROCESSOR

The general architecture of a reconfigurable RNS processor is shown in Figure 3. Given a moduli set hardware complexity depends on the functionalities of the RNS. Because of the space issue, a simplified structure is shown using only three arithmetic operators. It contains

    1. 2 Binary to RNS converters
    2. 7 MUXs
    3. Adder
    4. Subtractor
    5. Multiplier
    6. RNS to Binary converter [26][27]

Binary numbers are passed to the processor as inputs which first are converted to the RNS number. Here the selection of moduli is very much important because the proper selection of moduli optimizes the bit efficiency, area of the processor & time to process the particular function.

After the conversion of the binary number to its corresponding residue representation, the arithmetic operations can be performed. As any binary number produces a set of RNS numbers depending upon the number of moduli used, m copies of arithmetic units (adder, subtractor, multiplier etc) are required to perform some arithmetic operation of a number when it is converted to RNS, where m is the number of moduli used in that scheme. As the residues can be independently operated, parallel arithmetic operations can be performed on the residue set.

Figure 4 shows the control & data flows between the various paths. The Programmable Controller can program the RRNS Processor directly or Programmable Memory is used to store the bit stream. Programmable Controller is governed by the General purpose CPU.

In general, all modular arithmetic operations like Binary to RNS conversion or RNS addition, multiplication are implemented in chip by using two different methods [25]. One is the table look-ups, implemented by PLA. Second one is the Hybrid Methods, which is the combination of the legacy hardware, like full adders, with a table look-up, which can be used to convert the output of the legacy hardware to the correct residue format.

Use of PLA gives a faster hardware than the Hybrid method, but the later takes less area in the chip than the former. PLA can be a good choice of modern reconfigurable RNS processor because of their regular dense structure & easy interconnection.

The area & the speed of the reconfigurable RNS processor depends of the number of moduli in the moduli set, the method used for generation as well as the scheduling algorithm used.

## 6. DESIGN PROCEDURE

In the reconfigurable architecture, there is no fixed path between the device units, but the path can be changed depending upon the requirements. MUX are used before the inputs of the device units that act as the switch determining a specific path with respect to some particular select condition. For an example, suppose there are x number of adders, y number of subtractor & z number of multiplier in the processor. Also we are considering that the chip is accepting k number of inputs. So in general, for all the arithmetic unit having 2 inputs, the MUX in front of the inputs must be having of (x inputs coming from the outputs of adders + y inputs coming from the outputs of subtractors + z inputs coming from the outputs of multipliers + k external inputs). So the MUX must have $\lceil \log_2(x + y + z + k) \rceil$ select lines. In our example (Figure 3), for simplicity, we have taken x = y = z = 1, k = 2. So we can use 5 x 1 MUXes having 3 select lines before all the arithmetic devices in general. As the output coming from all the units are fed to the inputs of all the unit devices in general, it is possible to have any combination of the arithmetic operations computed by the processor.

In our work, the aim is to design a RNS processor which can be reconfigured dynamically to compute some pre-determined functions. For this, the unit operations need to be analysed & sequenced in terms of the inputs & arithmetic operations. As the arithmetic operations are depicted in terms of the select condition on the MUX, the inputs as well as the select conditions need to be stored using a LUT. The bit sequences are stored in the LUT block wise, each block has some particular address. When the address is given for some function, these inputs & select conditions are passed to the input MUXes.

Table 3. Comparison of area given by [28] for three, four, five six moduli set

| N | | $\{2^n, 2^n+1, 2^n-1, 2^n\pm1,...,2^n\pm1\}$ | | N | | $\{2^n, 2^n+1, 2^n-1, 2^n\pm1,...,2^n\pm1\}$ | |
|---|---|---|---|---|---|---|---|
| | | Moduli Set | Area | | | Moduli Set | Area |
| 12 | n=3 | (18,19,17) | 413 | 24 | n=3 | (256,257,255) | 898 |
| | n=4 | (8,9,7,11) | 281 | | n=4 | (64,65,63,67) | 593 |
| | n=5 | (6,7,5,11,13) | 354 | | n=5 | (28,29,27,31,25) | 499 |
| | n=6 | (4,5,3,7,11,13) | 322 | | n=6 | (16,17,15,19,23,11) | 471 |
| 16 | n=3 | (42,43,41) | 340 | 28 | n=3 | (646,647,645) | 3112 |
| | n=4 | (16,17,15,19) | 439 | | n=4 | (128,129,127,131) | 939 |
| | n=5 | (10,11,9,13,7) | 369 | | n=5 | (50,51,49,47,53) | 1019 |
| | n=6 | (8,9,7,11,5,13) | 426 | | n=6 | (26,27,25,29,23,31) | 621 |
| 20 | n=3 | (102,103,101) | 705 | 32 | n=3 | (1626,1627,1625) | 3292 |
| | n=4 | (32,33,31,35) | 523 | | n=4 | (256,257,255,259) | 1365 |
| | n=5 | (16,17,15,19,23) | 439 | | n=5 | (86,87,85,89,77) | 1113 |
| | n=6 | (12,13,11,17,7,19) | 463 | | n=6 | (42,43,41,47,37,53) | 928 |

Table 4. Comparison of area given by standard moduli schemes for three moduli set

| | Proposed Scheme | | $S_{M_1}=(2^n,2^n+1,2^n-1)$ | | $S_{M_2}=(2^n,2^n-1,2^{n+1}-1)$ | | $S_{M_3}=(2^{2n}+1,2^n+1,2^n-1)$ | |
|---|---|---|---|---|---|---|---|---|
| N | 3-Moduli Set | Area | 3-Moduli Set | Area | 3-Moduli Set | Area | 3-Moduli Set | Area |
| 6 | (6,7,5) | 156 | (8,9,7) | 191 | (8,7,3) | 169 | (17,5,3) | 186 |
| 10 | (12,13,11) | 268 | (16,17,15) | 305 | (16,15,7) | 276 | (65,9,7) | 304 |
| 16 | (42,43,41) | 340 | (64,65,63) | 442 | (64,63,31) | 431 | (257,17,15) | 433 |
| 24 | (256,257,255) | 898 | (512,513,511) | 1167 | (512,511,255) | 1089 | (4097,65,63) | 998 |
| 32 | (1626,1627,1625) | 3292 | (2048,2049,2047) | 4280 | (4096,4097,2047) | 4734 | (65537,257,255) | 4156 |

# 7. IMPLEMENTATION

The proposed scheme is implemented using a Virtex5 board (XC5VLX30). Verilog code is generated corresponding to the Reconfigurable RNS Processor which is synthesized & simulated using Xilinx.

Most of the types of equations containing the combination of arithmetic operations like addition, subtraction & multiplication can be implemented using the proposed scheme. Some examples are given here to illustrate the scheme.

Let the Function1 = $(X + Y) \times Z$

When the address of Function1 is given to the LUT, the corresponding block is activated. It will first supply the 2 binary numbers, X & Y, X to the BtoR Conv1 & Y to the BtoR Conv2.

Let m = 3, the number of moduli in the moduli set. So BtoR Conv1 & BtoR Conv2 will generate $(x_1, x_2, x_3)$ and $(y_1, y_2, y_3)$ respectively. Therefore,

SM1 = 000 & SM2 = 001 so that $(x_1, x_2, x_3)$ and $(y_1, y_2, y_3)$ can propagate to the inputs of the adder.

So TEMP1 = $(x_1 + y_1, x_2 + y_2, x_3 + y_3)$. In the next step Z must be supplied to the BtoR Conv2 and SM5 = 101 & SM6 = 001 so that $(x_1 + y_1, x_2 + y_2, x_3 + y_3)$ and $(z_1, z_2, z_3)$ can propagate to the inputs of the multiplier.

So TEMP2 = $((x_1 + y_1) \times z_1, (x_2 + y_2) \times z_2, (x_3 + y_3) \times z_3)$. Now set SM7 = 000 & it is enabled so that the output of the Multiplier passes to the RtoB Conv to have the final output of Function1 = $(X + Y) \times Z$ in binary.

Let the Function2 = X ^ Y

X & Y are fed to the BtoR Conv1 & BtoR Conv2 respectively to generate ($x_1$, $x_2$, $x_3$) and ($y_1$, $y_2$, $y_3$) respectively

Suppose POWER is the function unit (not shown in the Figure). POWER actually uses MULTIPLIER internally. When the select inputs of the corresponding MUXes are given, ($x_1$, $x_2$, $x_3$) are multiplied by itself ($y_1$, $y_2$, $y_3$) times using the same procedure shown before.

In Table 1 we compare area required for moduli set three, four, five, six generated by the moduli set $M_G$ is generated as $\{2^{n_1}, 2^{n_1}+1, 2^{n_1}-1, 2^{n_2} \pm 1, \ldots, 2^{n_t} \pm 1\}$ [28], where $n_1 > n_i$, and $n_1$ as well as $n_i's \ (i = 2,3,..,t)$ need to be chosen such that all these moduli are co-prime numbers. It may be noted that [28] is most widely accepted, hence the comparison three, four, five, six has been done. In Table 4 comparison has been done for three-moduli set with different schemes namely

$S_{M_1} = (2^n, 2^n + 1, 2^n - 1)$, $S_{M_2} = (2^n, 2^n - 1, 2^{n-1} - 1)$ and $S_{M_3} = (2^{2n} + 1, 2^n + 1, 2^n - 1)$. [18][28][29][30]

The results have been illustrated graphically in Figure 2. From these figure it is observed that the area of the Reconfigurable RNS Processor varies as the number of module.

Figure 2: Graphical representation of Comparison of area between standard approaches for three moduli set

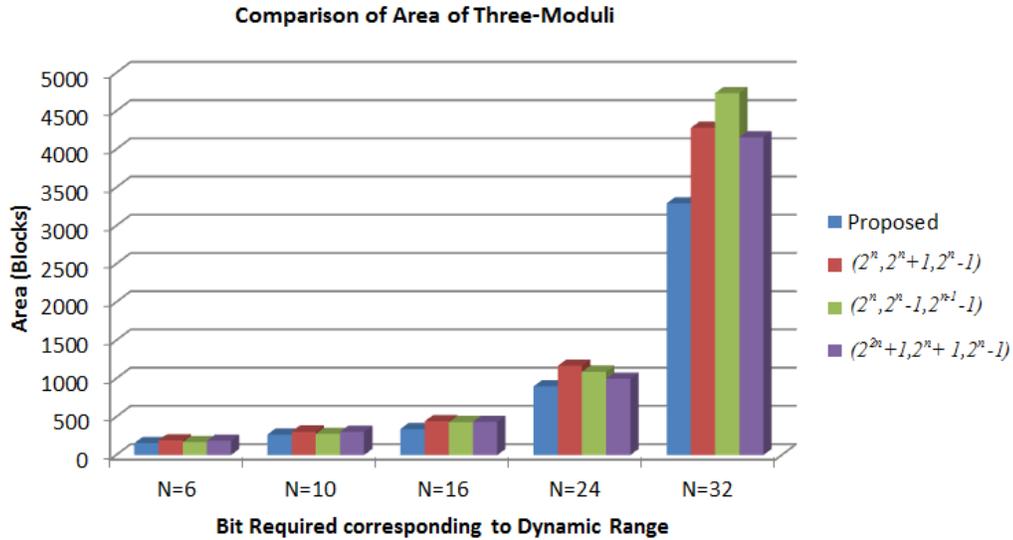

## 8. CONCLUSION

In this paper we proposed an algorithm to generate any moduli set of finite cardinality for a given dynamic range and given the proof of correctness for this proposed algorithm. We have also shown that bit efficiency of the proposed scheme is better than all other schemes given in the literature. In future we will be working on how these parameters, bit efficiency, h/w complexity and time can be optimized for a reconfigurable RNS processor. Another moduli set can be proposed which is better than our proposed scheme considering the three parameters mentioned above, using which we can get the optimized values of the same.

Figure 3: Simplified diagram of the proposed Reconfigurable RNS Processor

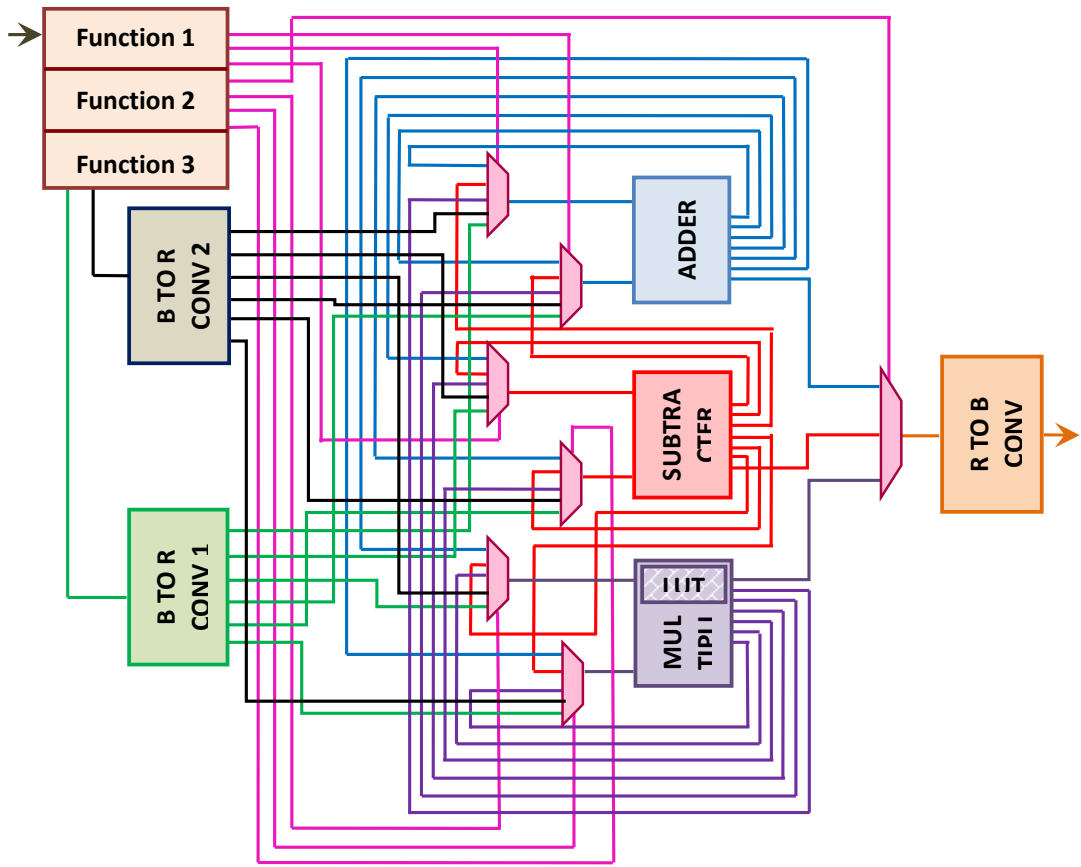

Figure 4: Control & Data flow in the proposed Reconfigurable RNS Processor

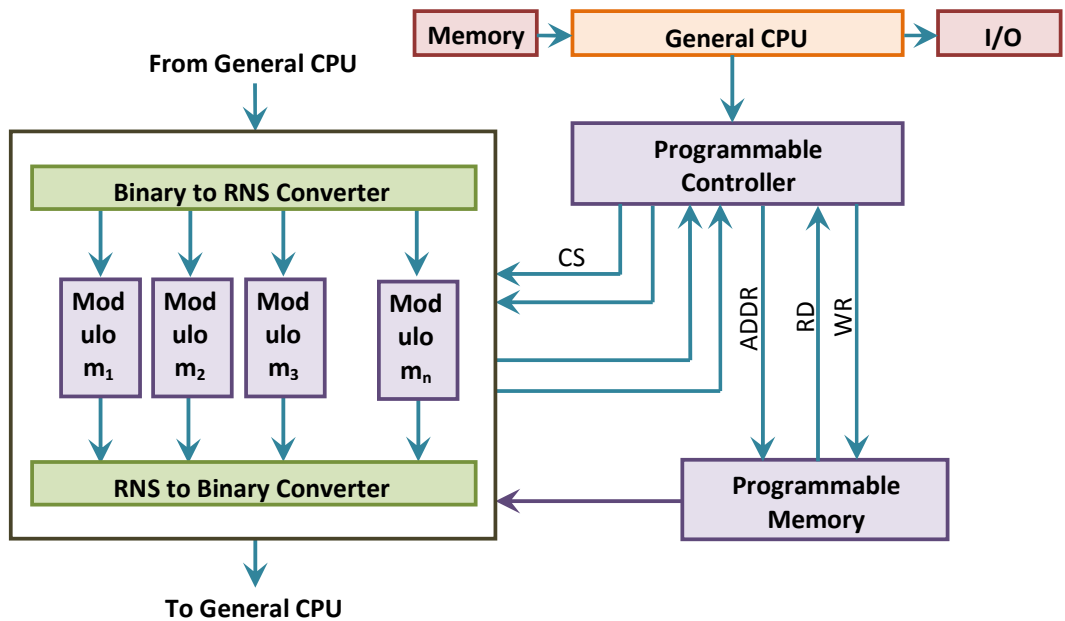

## Authors

**Chaitali Biswas Dutta** received the B.Sc. Degree in mathematics from University of Burdwan, India in 2003, the Master of Computer Application degree and MTech degree in Information Technology, both from West Bengal University of Technology, India, in 2006 and 2009 respectively. Currently she is a research scholar in the Department of Computer Science and Engineering, University of Kalyani, India. She is also an assistant professor, department of Computer Application, GIMT, Guwahati, India. Her research interest includes sensor network, wireless network security, VLSI, and so forth. She has published a few papers in international journals and conferences.

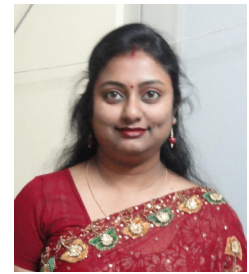

**Partha Garai** received the B.Sc. Degree in physics from University of Calcutta, India in 2002, the Master of Computer Application degree and MTech degree in Information Technology, both from West Bengal University of Technology, India, in 2006 and 2009 respectively. Currently he is a research scholar in the Machine Intelligence Unit, Indian Statistical Institute, Kolkata, India. His research interests include pattern recognition, machine learning, soft computing, VLSI, and so forth. He has published a few papers in international journals and conferences. He has received INSPIRE research fellowship for the years 2011-2016 from the Department of Science and Technology, Government of India.

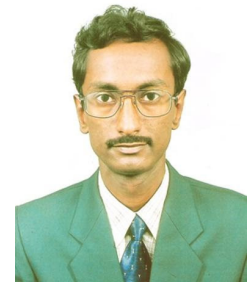

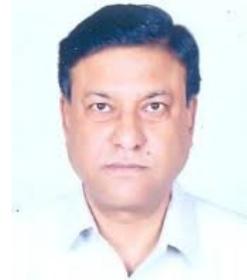

**Prof. Amitabha Sinha** obtained his Ph.D degree in Computer Science and Engineering from IIT, Delhi and has worked in senior positions in the industry and in academia for the last 25 years. He was Head, DSP (Digital Signal Processing) group of the R&D center of CMC Ltd. and Vice-President of HFCL. Prof. Sinha has taught at Oakland University, U.S.A., erstwhile B.E. College, Howrah (now BESU) and BITS, Pilani. He was a member of the advisory board of the Dept. of Science and Technology, Govt. of West Bengal. He is a member of IEEE and has chaired sessions of IEEE. His areas of expertise are embedded computer systems, application specific digital circuit design using FPGAs, DSP (Digital Signal Processing), parallel architecture and parallel processing for signal and imaging applications. He co-founded a U.S. based start up company "ESP microDesign" working on developing IPR in the area of Reconfigurable DSP Processor.